\documentstyle[prl,aps,preprint]{revtex}
\begin{document}
\tighten

\title{CHIRAL MULTIPLETS OF HADRON CURRENTS}
\author{{Thomas D. Cohen and Xiangdong Ji}
\thanks{On leave of absence for Department of Physics, MIT, Cambridge, MA.}}
\bigskip

\address{
Department of Physics, 
University of Maryland \\
College Park, Maryland 20742 \\
{~}}

\date{U. of MD PP\#97-066~~~DOE/ER/40762-106~~~ November 1996}

\maketitle

\begin{abstract}
Using group theoretical methods, we enumerate possible 
chiral representations in which hadron interpolating
currents can be classified. We give simple examples 
of currents in each representation, some of which are
well known. The classification enables one to 
find relations among current vacuum correlators 
in a chirally and/or axial U(1) symmetric phase of QCD in a 
straightforward way. Besides recovering many well-known 
relations among two-point correlation functions, 
a number of novel relations are found.

\end{abstract}

\pacs{xxxxxx}

\narrowtext

\section{Introduction}

Approximate chiral symmetry is an important
feature of the QCD lagrangian.  Much of the low-energy
behavior of QCD at zero temperature and density 
can be understood in terms of chiral symmetry, its
spontaneous breaking and the anomolous breaking of
the $U(1)_A$ subgroup.  There has been considerable 
interest in the possibility of a chirally
restored phase of QCD, as might be expected at 
a sufficiently high temperature, or a large number of flavors,  
or a high density \cite{shur1}. Recently, there have also 
been some speculations about the possibilities that 
$U(1)_A$ symmetry might be restored \cite{u1a1,u1a2,u1a3}, 
despite the fact that the anomaly 
in the singlet axial current is formally 
temperature-independent \cite{u1a}.
These issues are of obvious theoretical interest.  They
also may  have nontrivial experimental implications since
regions of the chirally restored phase might be produced
in ultrarelativistic heavy-ion collisions.

An important issue in attacking this problem theoretically
is finding calculable quantities which are sensitive
to whether the phase breaks the symmetry.
There are many well-known signatures of chiral symmetry
restoration; for instance, the vanishing of the 
chiral quark condensates and the disappearance of Goldstone
modes. A special class of tests
consists of comparing different thermal 
two-point correlation functions of currents with hadronic
quantum numbers;  in the symmetric phase they are connected
due to the underlying chiral (or possibly axial U(1)) symmetry.
We wish to observe that it is important in practice to check 
many of these signtures simultaneously since practical 
QCD-based calculations are limited to numerical calculations on the
lattice which necessarily have both statistical and systematic
errors.  Accordingly it is hard to tell whether the
approximate  vanishing of a single observable is an indication 
of symmetry restoration or simply an
accidently small value masked by numerical noise.   

Correlation functions with hadron quantum numbers 
are a useful window into the 
structure of excitations of the QCD vacuum. They 
have been used widely in the QCD sum rule 
and lattice calculations
to study hadron spectrum at 
zero temperature. One can construct an infinite
number of hadron currents of different dimensions
and Lorentz symmetry by employing covariant derivatives, 
gluon and quark fields. However, in  
realistic calculations, one  uses
simple  ones  with given quantum numbers. In the chirally-symmetric 
phase, chiral symmetry imposes
many relations among the current correlators; these 
are chiral Ward identities.
Some of these are well known and can be derived 
by simple inspections. 

The goal of this paper is to show 
that a very large class of hadron interpolating
currents fall into a few chiral representations 
and that relations among current correlators can be 
systematically derived using multiplications of these 
representations.  In particular, we will explictly enumerate
all of the chiral representations of 
mesonic currents that carry flavor quantum numbers of one quark
and one antiquark fields and the representations
of baryon currents that carry flavor
quantum numbers of three quark fields.  We
should note that this is a very large class of interpolating
currents---there are no restrictions on the number of covariant
derivatives, gluon fields, and quark pairs that are coupled to 
chiral singlets, and indeed no restriction that the 
current even be local (although they must be gauge invariant). 
While we have focused our attention to a restricted class 
of currents, the techniques used in this
paper can be extended straightforwardly to determine
the chiral representation for arbitrary currents,
such as the pion interpolating current of type $\bar q   \tau^a
 q \, \bar q i \gamma_5 q$. 
It is also worth noting that
virtually all practical lattice gauge or QCD
sum rule calculations have used currents in the class considered here. 

The role of the chiral multiplet structures of the
currents on two-point correlation functions in the chirally
restored phase is  significant in two ways.  The first
concerns correlation functions between two distinct 
currents with the same  flavor,
spin, and parity quantum numbers but which 
belong to distinct chiral representations.  In a chirally-broken 
phase such as the $T=0$ vacuum, such correlation 
functions are generically nonzero.  In terms of a 
$T=0$ spectral representation of the correlator, this
indicates nothing more than the fact that each of these
currents has a nonzero overlap between the vacuum and the
same physical state.  However in a chirally restored phase
all such mixed correlators are identically zero. This
vanishing of all these mixed correlators can be used as a 
signature of symmetry restoration. Some examples 
have already been considered in Ref. \cite{shaf}.
When one enumerates the chiral representations one sees
that for virtually every flavor, spin and parity
channel there are interpolating currents with at least two
distinct sets of  chiral quantum numbers, even when restricting
to the class of currents considered here. 
This is significant when trying to interpret the nature of
the chirally
restored phase.  For example, questions such as ``does the $\rho$
meson survive the chiral transition?'' becomes intriniscally
ambiguous.  The question cannot even be formulated  without
specifing the chiral quantum numbers of the current coupling to
the $\rho$ channel.

The second class of issues concerns the equality of
certain two-point correlation functions.  If a current is in 
a nontrivial chiral multiplet, then under chiral rotations
one generates new currents with distinct  parity and/or flavor
quantum numbers.  Clearly, in a chiral restored phase these
newly generated currents must yield correlation functions
identical to the original ones and hence one predicts that
certain correlators must be identical.  Of course, many such 
examples have long been known. For example, for two massless
flavors the correlators in
the $\sigma$ (scalar-isoscalar) and $\pi$
(pseudoscalar-isovector) channels
corrresponding to the currents, $\bar q\tau^a q$ and $\bar q \tau^a\, i
\gamma_5  q$, respectively are  well known to be the same in the restored phase;
similary the $\rho$ (vector-isovector) and $A_1$
(pseudovector-isovector) corresponding to the currents $\bar q
\gamma^i \tau_a  q$ and $\bar q \gamma_5 \gamma^i
\tau_a q$ are also well known to be identical. 
However, there are many examples which are less familiar.
For example, the $\rho$ and the $b_1$ (psuedovector-isovector and
charge conjugation {\it odd}) corresponding to the currents
$ \bar q \tau^a \sigma^{ij} q$ and
$\bar q \tau^a \sigma^{0i} q$, respectively, are also
identical in this phase.  We note that there is a very large
number of these relations including a certain nucleon
interpolating current whose correlator in the chirally restored
phase is identical to correlators in the $\Delta \,
({\frac{1}{2}}^{-})$ channel.  

It is also interesting to  
consider correlator relations which would result  
if the  $U(1)_A$ symmetry were to be restored. 
Whether  the effects of the $U(1)_A$ anomaly play 
a role in the  chirally restored phase and if they
do how do, these effects die off with increasing 
temperature, remain interesting questions.  Accordingly,
it is useful to classify good signatures of the effects
of a manifest $U(1)_A$ symmetric phase on correlation
functions.  In the course
of our discussion, we will recover 
many relations which are widely known and have 
already been used frequently in the literature. However, besides 
organizing those into categories, we also
find a number of new relations which we suggest
will provide further insights 
for lattice or model calculations.

We divide our dicussions into meson and baryon
currents. For each case, we consider the possibilities
of two  and three massless flavors separately.
Of course, in nature there are neither two nor three
flavors of massless quarks.  For the analysis here
to be of any use in connecting to real QCD, it is important
that for quantities of interest the quark masses must 
be small enough to either neglect outright or to include
perturbatively in some type of chiral perturbation theory.
The up and down quark masses are presumably light enough 
for such a procedure to make sense.  Moreover, the nature
of the chiral symmetry of the underlying theory allows one
to deduce certain relations between correlators which hold
to order $m_q^2$ rather than $m_q$, thereby enhancing the
range of validity of the chiral expansion\cite{BCM2}.
Treating the strange quark mass as small is clearly far more
problematic.  It is by no means clear that a chiral expansion
in $m_s$ will be valid for any given quantity of interest
particularly in the vicinity of the phase transition. 
Even if it turns out that an expansion in $m_s$ is not valid,
the three massless flavor results are not 
without interest.  One obvious use is in providing limiting
cases against which to test more realistic calculations.  For
example, lattice calculations which hope to connect to the
real world of two light and one intermediate mass flavors
could be re-run comparatively easily with three light flavors.
The ability of such calculations to reproduce the correlator
relations for three massless flavors will be quite useful in
demonstrating that the systematic and statistical errors
inherent in the calculations are under control.

\section{meson currents}

The simplest meson currents can be constructed 
from one-quark and one-hermitian-conjugated-quark 
fields. Although one can construct more
complicated meson currents by including 
covariant derivatives, gluon and quark fields,
a large class of meson
currents belong to the chiral multiplets
of one-quark and one-antiquark product
representations. For $N_f$ flavors, 
the relevant chiral multiplets 
are representations 
of $SU(N_f)_L \times SU(N_f)_R$. 
In the following subsections, we consider 
the simplest chiral multiplets for 
two and three massless flavors.

\subsection{Two Massless Flavors}

The quark fields $(u_L, d_L)$ and
$(u_R, d_R)$ belong to the basic
representation $({1\over 2}, 0)+ (0,{1\over 2})$
of $SU(2)_L\times SU(2)_R$. 
The convention for denoting chiral multiplets is that 
the first and second numbers in a bracket refer 
to $SU(2)_L$ and $SU(2)_R$ representations, 
respectively. Under parity
transformation, the left-handed fields become 
right-handed and vice versa. In the case of $SU(2)$, 
the hermitian-conjugation fields transform according to
the same representation as the original fields. 

To classify meson currents, we consider the 
product representation, $[({1\over 2}, 0)+ (0,{1\over 2})] 
\times [({1\over 2}, 0) + (0,{1\over 2})]$. 
The simple angular momentum addition rules yield,
\begin{eqnarray}
&& \hspace*{-.5in}\left[({1\over 2}, 0)+ (0,{1\over 2})\right] \times \left[({1\over 2}, 0)
+ (0,{1\over 2})\right]  \nonumber \\
= && \Big[(\tilde 0,0)+(0,\tilde 0)\Big] + \left[({1\over 2}, {1\over 2})+
({1\over 2}, {1\over 2})\right] + \Big[(1,0)+(0,1)\Big] \ , 
\end{eqnarray}
where tilde on $\tilde 0$ has no group-theoretical role.
It simply serves as
 a reminder that this singlet corresponds a  
left- or right-handed quark-antiquark pair  coupled into 
a flavor-singlet rather than the absence of any fields. We use square brackets to group together
representions that are parity-conjugates.

Let us first consider left and right singlets $(\tilde 0,0)+(0,\tilde 0)$.
The correponding quark bilinears are $\bar u_Lu_L+\bar d_Ld_L$
and $\bar u_Ru_R+\bar d_Rd_R$. To form currents with good parity, 
we have to consider symmetric and antisymmetric combinations:
$(\bar u_Lu_L+\bar d_Ld_L) \pm (\bar u_Ru_R+\bar d_Rd_R)$. 
The simplest currents with these flavor structures 
have the quantum numbers of $\omega(1^{--})$ and $f_1(1^{++})$, 
\begin{eqnarray}
       j^\mu_{vI=0} &=& \bar q\gamma^\mu q\ ,  \nonumber \\
      j^{\mu}_{aI=0} &=& \bar q\gamma^\mu \gamma_5 q \ , 
\label{2}
\end{eqnarray}
where $q$ is a column vector consisting of up and down quark fields.
Both currents are invariant under $SU(2)_L\times SU(2)_R$ 
and $U(1)_A$ transformations. Therefore, flavor symmetries impose 
no constraints on their correlation functions. 
 
Next, we consider $({1\over 2}, {1\over 2})+
({1\over 2}, {1\over 2})$. Under the 
isospin subgroup, they contain both isoscalar and isovector
multiplets. The isoscalar ones with good parity 
correspond to the quark bilinears
$(\bar u_Lu_R + \bar d_Ld_R) \pm(\bar u_Ru_L+ \bar d_Rd_L)$.
Two examples of the positive parity currents are,
\begin{equation}
       j_{sI=0} = {1\over 2} \bar qq \ , ~~~~ 
       j_{tI=0}^k= {1\over 4}\epsilon^{ijk} \bar q \sigma^{ij} q \ , 
\end{equation}
which have the quantum numbers of $\sigma(0^{++})$ and 
$h_1(1^{+-})$, respectively.
And the corresponding examples of the negative parity currents are, 
\begin{equation}
       j_{pI=0} = {1\over 2} \bar qi\gamma_5q, ~~~~ 
        j_{\tilde tI=0}^k= {1\over 2} \bar q\sigma^{0k} q \ , 
\end{equation}
which have the quantum numbers of $\eta(0^{-+})$
 and $\omega(1^{--})$, 
respectively. 
Examples of isovector multiplets 
can be obtained by inserting the isospin Pauli matrices 
$\tau^a (a=1,2,3)$ into the above currents, 
\begin{eqnarray}
        &&j_{sI=1}^a = \bar q{\tau^a\over 2} q \ ;  ~~~~ j_{\tilde tI=1}^{ka} 
        = {1\over 2}\epsilon^{ijk}\bar q {\tau^a\over 2}\sigma^{ij} q \ ; 
\nonumber \\
        &&j_{pI=1}^a = \bar q{\tau^a\over 2}i\gamma_5q \ ; ~~~~ 
       j_{tI=1}^{ka} = \bar q {\tau^a\over 2}\sigma^{0k} q \ . 
\end{eqnarray}
which have the quantum numbers of $\delta(0^{++})$, 
$b_1(1^{+-})$, $\pi(0^{-+})$, and
$\rho(1^{--})$, respectively. 

Under $SU(2)_L\times SU(2)_R$, the isoscalar and isovector currents
transform into each other in pairs,
\begin{eqnarray}
          j_{sI=0}~~  && \leftrightarrow ~~j_{pI=1}^a \ , 
    \nonumber \\
          j_{pI=0}~~  && \leftrightarrow ~~j_{sI=1}^a \ , 
     \nonumber \\
          j_{tI=0}^k~~ && \leftrightarrow ~~j_{\tilde tI=1}^{ka} \ ,        
    \nonumber \\ 
          j_{\tilde t I=0}^k~~ && \leftrightarrow ~~j_{tI=1}^{ka} \ . 
\end{eqnarray}
Thus, in the chirally-symmetric phase, 
the bilocal current correlators of each pair
are the same, 
\begin{eqnarray}
      \langle  Tj_{sI=0}(x)j_{sI=0}(0)\rangle
    &=& \langle  Tj_{pI=1}^a(x)j_{pI=1}^a(0)\rangle \ ,  \nonumber \\ 
      \langle Tj_{pI=0}(x)j_{pI=0}(0)\rangle
    &=& \langle Tj_{sI=1}^a(x)j_{sI=1}^a(0)\rangle \ ,   \nonumber \\ 
      \langle Tj_{tI=0}^k(x)j_{tI=0}^k(0)\rangle
    &=& \langle Tj_{\tilde tI=1}^{ka}(x)j_{\tilde tI=1}^{ka}(0)\rangle \ ,   
\nonumber \\ 
      \langle Tj_{\tilde tI=0}^k(x)j_{\tilde tI=0}^k(0)\rangle
    &=& \langle Tj_{tI=1}^{ka}(x)j_{tI=1}^{ka}(0)\rangle  \ . 
\end{eqnarray}
Notice that there is no summation for repeated indices in the above 
equation. Unless stated explicitly, the same is 
used for other equations below. 
The first two relations say that the $\pi$ ($\delta$)
type of correlators are the same as
$\sigma$ ($\eta$) type of correlators, a result which is familiar. 
The second two relations say that 
the $\rho$ ($b_1$) type of correlators are the same as $h_1$ 
($\omega$) type, which is lessknown.  

On the other hand, under $U(1)_A$ transformations, 
the currents with opposite parities transform into
each other. If $U(1)_A$ symmetry is restored in some phase, 
 we then have the following relations among the
correlators,
\begin{eqnarray}
      \langle Tj_{sI=0}(x)j_{sI=0}(0)\rangle
    &=& \langle Tj_{pI=0}(x)j_{pI=0}(0)\rangle\ ,    \nonumber \\ 
      \langle Tj_{sI=1}^a(x)j_{sI=1}^a(0)\rangle
    &=& \langle Tj_{pI=1}^a(x)j_{pI=1}^a(0)\rangle\ ,    \nonumber \\ 
      \langle Tj_{tI=0}^k(x)j_{tI=0}^k(0)\rangle
    &=& \langle Tj_{\tilde tI=0}^k(x)j_{\tilde tI=0}^k(0)\rangle\ ,    
          \nonumber \\ 
      \langle Tj_{ tI=1}^{ka}(x)j_{tI=1}^{ka}(0)\rangle
    &=& \langle Tj_{\tilde tI=1}^{ka}(x)j_{\tilde tI=1}^{ka}(0)\rangle \ .   
\end{eqnarray}
The isovector relations are particularly interesting because
they contain no disconnected contributions in the path-integral 
formulation. In the literature,
the $\pi$ and $\delta$ types of correlators have been compared at
the chiral transition region to learn about the fate of $U(1)_A$
symmetry\cite{shur1,shaf,columbia,milc}. The same comparison can be made
of the $\rho$ and $b_1$ types of correlators.

Finally, we consider $(0,1)+(0,1)$ which contain
isovector multiplets only. 
The simplest example of the multiplets is,
\begin{eqnarray}
      j^{\mu a}_{vI=1} & = &  \bar q \gamma^\mu {\tau^a\over 2} q \ , \nonumber \\
      j^{\mu a}_{aI=1} & = & \bar q \gamma^\mu \gamma_5{\tau^a\over 2} q \ , 
\end{eqnarray}
which have the quantum numbers of $ \rho$ 
and  $A_1$, respectively.  It is worth noting that the currents
written above  are both conserved in the massless limit and
hence do  couple only to vectors (axial vectors) and not
to scalars (pseudoscalars).  More general realization of
currents in this representation such as 
\begin{eqnarray}
      j^{\mu a}_{vI=1} & = &  \bar q \gamma^\mu {\tau^a\over 2}
      F^2 q \ , \nonumber \\
      j^{\mu a}_{aI=1} & = & \bar q \gamma^\mu \gamma_5{\tau^a\over 2}
      F^2 q \ , 
\end{eqnarray}
where $F^2=F^{\alpha\beta}F_{\alpha\beta}$, 
have the quantum numbers  of $(\delta,\rho)$ and $(\pi,A_1)$, 
respectively.

Under $U(1)_A$ transformations, the currents are invariant.
Under $SU(2)_L\times SU(2)_R$ 
chiral transformations, they mix with each other.
Thus if the vacuum is chirally-symmetric,
their two-point correlation functions are equal;
\begin{eqnarray}
      \langle Tj_{vI=1}^{\mu a}(x)j_{vI=1}^{\mu a}(0)\rangle
    = \langle Tj_{aI=1}^{\mu b}(x)j_{aI=1}^{\mu b}(0)\rangle \ ,  
\end{eqnarray}
which is a well-known result.

A slightly more complicated example of (0,1)+(1,0) multiplet
involves currents with gluon fields,
\begin{eqnarray}
     \tilde j^{\mu a}_{vI=1} & = &  \bar q \gamma^\nu F_{\mu\nu} {\tau^a \over 2}
 q \  , \nonumber \\
     \tilde j^{\mu a}_{aI=1} & = & \bar q \gamma^\nu \gamma_5 
       F_{\mu\nu} {\tau^a\over 2} q \ . 
\end{eqnarray}
The first current has the quantum number of
the exotic vector meson $1^{-+}$
and the second has that of $b_1(1^{+-})$. The chiral symmetry
again predicts the equality of the two-point correlators in 
the symmetric phase.

\subsection{Three Massless Flavors}

For three massless flavors, the meson currents belong
to the product representations of $(3,1)+(1,3)$ and $(\bar 3, 1)
+(1, \bar 3)$. The former correspond to the quark fields $u_L,d_L,s_L$
and $u_R,d_R,s_R$; the latter correspond to the
the conjugate quark fields $\bar u_L, \bar d_L, \bar s_L$ 
and $\bar u_R, \bar d_R, \bar s_R$.  SU(3) multiplication rules
give, 
\begin{eqnarray}
&& \hspace*{-.5in}\Big[(\bar 3, 1)+ (1,\bar 3)\Big] \times \Big[(3, 1)
+ (1,3)\Big]  \nonumber \\
= && \Big[(\tilde 1,1)+(1,\tilde 1)\Big] + \Big[(\bar 3, 3)+
(3, \bar 3)\Big] + \Big[(8,1)+(1,8)\Big] \ . 
\end{eqnarray}
The notations for SU(3) representations are such that
they denote the actual dimensions.

Let us consider first the chiral-singlet 
$(\tilde 1,1)+(1,\tilde 1)$. One can easily construct
currents that are generalizations
of those in $(\tilde 0,0)+(0,\tilde 0)$ of the two-flavor case
(Eq.(\ref{2})), 
\begin{eqnarray}
        j^{\mu}_{v1} & = & \bar q \gamma_\mu q \ , \nonumber \\
        j^{\mu}_{a1} & = & \bar q \gamma_\mu\gamma_5 q \ , 
\end{eqnarray}
where $q$ is now a column vector consisting of up, down, and
strange quark fields. Both $j^{\mu}_{v1}$ and $j^{\mu}_{a1}$
are invariant under $SU(3)_L\times SU(3)_R$ and 
$U(1)_A$ transformations, and therefore chiral symmetries
do not impose any constraints on their correlators. 

The multiplet $(\bar 3, 3)+(3, \bar 3)$ 
contains $SU(3)_V$ octets and singlets. 
The eighteen $J=0$ currents constructed from 1 and 
$\gamma_5$ matrices belong to this chiral multiplet,
\begin{eqnarray}
       j_{s1} & = &  \bar q q/\sqrt{6} \ ,      \nonumber \\ 
       j_{p8}^a & = & \bar q i\gamma_5 t^a q \ ,  \nonumber \\
       j_{p1} & = & \bar q i\gamma_5 q/\sqrt{6}  \ ,  \nonumber \\
       j_{s8}^a & = & \bar q t^a q \ ,  
\end{eqnarray}
where $t^a=\lambda^a/2$ and $\lambda^a ~(a=1,...,8)$ are Gell-Mann
matrices. Under $SU(3)_L\times SU(3)_R$, these currents
transform into each other, and their two-point correlators 
equal in a chirally symmetric phase, 
\begin{equation}
     \langle Tj_{s1}(x)j_{s1}(0)\rangle = 
     \langle Tj_{s8}^a(x)j_{s8}^a(0)\rangle =
     \langle Tj_{p1}(x)j_{p1}(0)\rangle = 
     \langle Tj_{p8}^a(x)j_{p8}^a(0)\rangle \ . 
\label{j0}
\end{equation}
Using the currents in this class one sees that the correlator 
in the pion channel  is necessarily the same as  that in the
$\eta^\prime$ channel despite the existence of the anomaly. 
This result has previously been obtained from arguments 
based on instanton contributions\cite{eta1,eta2} and explictly 
on the basis of group theory \cite{birse}. 

One can also construct eighteen $J=1$
currents in the same chiral multiplet 
from the $\sigma^{\mu\nu}$ matrix, 
\begin{eqnarray}
       j_{t1}^k & = &  \bar q \sigma^{0k} q/\sqrt{6} \ ,      \nonumber \\ 
       j_{\tilde t8}^{ka} & = & {1\over 2}\epsilon^{ijk}\bar q 
               \sigma^{ij} t^a q \ ,   \nonumber \\
       j_{\tilde t1}^k & = & {1\over 2}\epsilon^{ijk}\bar q 
               \sigma^{ij} q/\sqrt{6}  \ ,  \nonumber \\
       j_{t8}^{ka} & = & \bar q \sigma^{0k} t^a q \ , 
\end{eqnarray} 
which have the quantum numbers of $J^{\pi C}=1^{\pm+}$ mesons:
$\rho, \omega, \phi, K^*, b_1, h_1, K_{1B}$. 
Under $SU(3)_L\times SU(3)_R$, these currents 
transfom into each other in the same way as 
the $J=0$ multiplet does. In the chirally-symmetric
phase, their two-point correlators have similar
relations as those in Eq. (\ref{j0}),
\begin{equation}
     \langle Tj_{t1}^k(x)j_{t1}^k(0)\rangle = 
     \langle Tj_{t8}^{ka}(x)j_{t8}^{ka}(0)\rangle =
     \langle Tj_{\tilde t1}^k(x)j_{\tilde t1}^k(0)\rangle = 
     \langle Tj_{\tilde t8}^{ka}(x)j_{\tilde t8}^{ka}(0)\rangle \ . 
\end{equation} 

Finally, we consider the multiplet $(1,8)+(8,1)$ which contains
$SU(3)$ flavor octets. The simplest currents in the multiplet are, 
\begin{eqnarray}
        j_{v8}^{\mu a} &=& \bar q \gamma_\mu t^a q \ , \nonumber \\
        j_{a8}^{\mu a} &=& \bar q \gamma_\mu \gamma_5 t^a q \ . 
\end{eqnarray}
Under $SU(3)_L\times SU(3)_R$, they transform into each other
and thus their two-point correlators equal in the chirally-symmetric 
phase, 
\begin{eqnarray}
      \langle Tj_{v8}^{\mu a}(x)j_{v8}^{\mu a}(0)\rangle
    = \langle Tj_{a8}^{\mu b}(x)j_{a8}^{\mu b}(0)\rangle \ ,  
\end{eqnarray}
Under $U(1)_A$ tranformations, the currents 
are separately invariant.

Since there are no two simple currents  
which belong to the same chiral mutiplet
but transform differently under the $U(1)_A$ group, 
one cannot form simple two-point correlators to
test the $U(1)_A$ restoration, as in the two-flavor case. 
One can, however, construct 
three-point correlators that are chiral-singlet,
but transform nontrivially under $U(1)_A$. An example
is presented in Ref. \cite{birse}. 
 
\section{Baryon currents}

Assuming color SU(3) symmetry, one can construct the simplest 
baryon currents out of three quark fields. However, a large
class of baryon currents can be classified in the
chiral multiplets derived from 
the product of three basic (quark) representations.
In the following subsections, we again consider two and 
three massless flavors separately.
 
\subsection{Two Massless Flavors}

For two massless flavors, we consider baryon
currents belonging to $[(0,{1\over 2})+({1\over 2},0)]^3$. 
Reducing it to irreducible multiplets, 
we find, 
\begin{eqnarray}
     \left[(0,{1\over 2})+({1\over 2},0)\right]^3 &
       = &\left[({3\over 2},0)+(0,{3\over 2})\right]
           +3\times \left[(1,{1\over 2})+({1\over 2},1)\right]  \nonumber \\
  && +3\times \left[(\tilde 0,{1\over 2})
       + ({1\over 2}, \tilde 0)\right] + 2\times \left[(\tilde {1\over 2},0)+(0,
      \tilde {1\over 2})\right]\ . 
\end{eqnarray}
Multiple appearances of the same representations
are due to the permutation symmetry of 
three quark labellings. The tildes on 
$\tilde 0$ and $\tilde {1\over 2}$ serve as a reminder that 
a pair of left- or right-handed quarks has been coupled to 
flavor-singlet.     

One of the simplest examples of currents 
in multiplet $({1\over 2}, \tilde 0)+( 
{1\over 2}, \tilde 0)$, is the 
spin-1/2 proton interpolating field,
\begin{equation}
    \eta_{N} = \left(u^TC\gamma_\alpha u\right) \gamma_5 \gamma^\alpha d \ , 
\label{19}
\end{equation}
where and hereafter 
color indices on the quark fields are implicit
and totally antisymmetric. This current 
has been used in the QCD sum rule
calculations \cite{ioffe}. The current itself 
couples also to a negative-parity spin-1/2 state, 
\begin{equation}
     \langle 0|\eta_N|N({1\over 2})^-\rangle = \lambda_N'\gamma_5 U(p) \ ,
\end{equation}
where $U(p)$ is a Dirac spinor. [For a recent
application of this feature, see Ref. \cite{oka}.]
Consider the following two-point correlator:
\begin{equation}
   \int d^4x e^{ixp}\langle T \eta_{N}(0)\bar \eta_{N}(x)\rangle
       = \rho_1(p^2)  p_\mu \gamma^\mu + \rho_2(p^2)\ . 
\end{equation}
In the chirally-symmetric phase, by making chiral
rotation $U=\exp(i\pi \tau^3 \gamma_5/2)$, one can show,
\begin{equation}
       \langle T \eta_{N}(0)\bar \eta_{N}(x) \rangle=
       -\gamma_5 \langle T \eta_{N}(0)\bar \eta_{N}(x)\rangle \gamma_5 \ . 
\end{equation}
Thus, we have $\rho_2(p^2)=0$, i.e. 
the correlator contains only the chiral-even term.
Since negative parity states contribute 
to $\rho_2(p^2)$ with an opposite sign compared with 
positve parity states, the above result
implies that every intermediate state has a degenerate 
partner of opposite parity and their chiral-odd spectral
strengths cancel. As we shall see below, this is quite a 
general property of two-point baryon correlators in 
the chirally-symmetric phase if the chiral limit
can be taken uniformly.

The simplest current in the 
$(\tilde{1\over 2}, 0)+(0, \tilde{1\over 2})$ multiplet is, 
\begin{equation}
    \eta_{N'} = \left( u^TC\sigma_{\alpha\beta} u\right) 
     \gamma_5 \sigma^{\alpha\beta}d \ , 
\label{22}
\end{equation}
which has also been recognized in the QCD
sum rule calculations\cite{ioffe}. Again the two-point
correlator has only the chiral-even term
if the vacuum has chiral symmetry. 

>From the  group theoretical standpoint, the 
two multiplets discussed so far are identical.
Thus, their product can produce an 
$SU(2)_L\times SU(2)_R$ singlet, and
the correlation function,
\begin{equation}
   \int d^4x e^{ixp}\langle T \eta_{N}(0)\bar \eta_{N'}(x)\rangle\ , 
\end{equation}
is nonzero even if the vacuum is chirally symmetric.
[Of course, in that case the $\rho_2(p^2)$ type of term does
vanish.] However, since $\eta_{N}$ and $\eta_{N'}$
transform differently under $U(1)_A$, the chiral-even
term would vanish if the $U(1)_A$ symmetry is restored.
This interesting diagnosis for $U(1)_A$ restoration
signature was first studied by Schafer and Shuryak
\cite{shaf}. 

The chiral multiplet $({1\over 2},1)+(1, {1\over 2})$ 
contains both $I={1\over 2}$ and $I={3\over 2}$ isospin multiplets. 
The simplest $I={1\over 2}, I_z={1\over 2}$ current is \cite{ioffe}, 
\begin{equation}
       \eta_{N}^\mu = \left(u^T C \sigma_{\alpha\beta} u \right)\gamma_5
            \sigma^{\alpha\beta}\gamma^\mu d - 
 \left(u^TC \sigma_{\alpha\beta} d \right) \gamma_5
            \sigma^{\alpha\beta}\gamma^\mu u \ ,
\label{24}
\end{equation}
which has the quantum numbers of the nucleon($P_{11}$), as well as $S_{11}$, 
$P_{13}$ and $D_{13}$ resonances. 
Under $SU(2)$ chiral rotation, for instance $U=\exp(i\pi\tau^3\gamma_5/4)$,
the current is tranformed to its $I={3\over 2}$ partner, 
\begin{equation}
     \eta_{\Delta}^\mu = \left(u^T C \sigma_{\alpha\beta} u\right) \gamma_5
            \sigma^{\alpha\beta} \gamma^\mu d  +
          \left(u^T C \sigma_{\alpha\beta} d\right) \gamma_5
            \sigma^{\alpha\beta}\gamma^\mu u \ , 
\label{25}
\end{equation}
which has the quantum numbers of $\Delta (P_{33})$, $D_{33}$, 
$S_{31}$, and $P_{31}$ resonances. A slightly different form
of $\eta^\mu_\Delta$ with three up-quark fields was first
used in a QCD sum rule calculation \cite{ioffe}.
In a chirally-symmetric phase, 
\begin{equation}
    \langle T \eta_{N}^\mu(x)\bar \eta_{N}^\nu(0) \rangle 
      =    \langle T \eta_{\Delta}^\mu(x)\bar \eta_{\Delta}^\nu(0) 
      \rangle \ , 
\end{equation}
and only chiral-even terms contribute.
If relevant baryons survive the chiral phase transition
and they couple to these currents strongly, 
$J={1\over 2}^\pm ({3\over 2}^\pm) $, $I={1\over 2}$ 
resonances would be degenerate with 
 $J={1\over 2}^\pm ({3\over 2}^\pm) $, $I={3\over 2}$ resonances. 

Finally, the chiral multiplet $({3\over 2},0)+(0,{3\over 2})$ 
has isospin ${3\over 2}$. The simplest current in this multiplet is,
\begin{equation}
    \eta_{\Delta}^{\mu\nu} = \left[\left( q^TC
     \sigma_{\alpha\beta} q\right)\gamma_5\sigma^{\alpha\beta}
      \sigma^{\mu\nu} q \right]_{I=3/2} \ , 
\end{equation}
where the flavor indices are coupled in a totally-symmetric way. 
$\eta_{\Delta}^{\mu\nu}$ can couple to $J={1\over 2}^\pm, 
{3\over 2}^\pm$ resonances. We haven't found any previous use of 
this current in the literature.  
In a chirally-symmetric phase,
the two-point correlator of the current contains chiral-even terms
only, 
which implies that parity partners are degenerate and the chiral-odd
spectral strengths cancel.

\subsection{Three Massless Flavors}

To classify baryon currents in three massless flavors, we 
consider decomposition of the representation 
$[(1,3)+(3,1)]^3$ of $SU(3)_L\times SU(3)_R$. 
The $SU(3)$ multiplication rules yield,
\begin{eqnarray}
     \Big[(1,3)+(3,1)\Big]^3 &= &\Big[(10,1)+(1,10)\Big]
           +3\times \Big[(6,3)+(3,6)\Big]  \nonumber \\
  & +& 3\times \Big[(\bar 3,3)
       + (3, \bar 3)\Big] + 2\times \Big[(8,1)+(1,8)\Big] + 
    \Big[(\tilde 1,1)+(1,\tilde 1)\Big] \ ,  
\end{eqnarray} 
where $\bar 3$ is from an antisymmetric combination of
two quark fields, and  $\tilde 1$ from an antisymmetric 
combination of three quark fields.
    
The chiral multiplet $(8,1)+(1,8)$ contains $SU(3)_V$
flavor octets.
Currents in this representation 
can be constructed as generalizations of the
currents in $(\tilde {1\over 2},0)+(0,\tilde {1\over 2})$ of the 
two-flavor case.
For instance, an extension of the $J={1\over 2}^{\pm}$ currents
in Eq. (\ref{22}) is ,
\begin{equation}
       \eta^a_{[8]} = \left[\left(q^TC\sigma_{\alpha\beta} q \right)\gamma_5
         \sigma^{\alpha\beta}q\right]_{[8]}^a \ ,
\end{equation} 
where the first two quark fields are symmetrized in flavor indices
to form a 6 of $SU(3)_V$. The complete flavor-octet wave functions 
can be found, for instance, in Ref. \cite{close}. In a
chirally-symmetric phase, the correlators
$\langle T \eta^a_{[8]}(x)\eta^a_{[8]}(0)\rangle$
contain chiral-even Dirac structure only. 

The multiplet $(3,\bar 3)+(\bar 3,3)$ contain both
$SU(3)_V$ octets and singlets. The baryon currents
of the multiplet can be obtained from generalizations of 
the currents in $({1\over 2},\tilde 0)+(\tilde 0,
{1\over 2})$ of the two-flavor case. For instance, from 
Eq. (\ref{19}) we can write down nine $J={1\over 2}^{\pm}$ currents,
\begin{eqnarray}
         \eta^a_{[8]'} &= &\left[\left(q^T C\gamma_\alpha \gamma_5 q\right) 
              \gamma^\alpha q\right]_{[8]'}^a 
      \ , \nonumber \\
       \eta_{[1]} &=& \left[\left(q^T C\gamma_\alpha \gamma_5 q\right) 
              \gamma^\alpha q\right]_{[1]} \ . 
\end{eqnarray}
where the first two quark fields are antisymmetrized in flavor
indices to form a $\bar 3$ of $SU(3)$. The complete flavor 
wave functions again can be found in Ref. \cite{close}. 
$\eta^a_{[8]'}$ have been used to calculate masses of 
the baryon-octet in the QCD sum rule approach in Ref. \cite{yaz}. 
The $\eta_{[1]}$ current carries the quantum number of 
a singlet $\Lambda$.
In the chirally-symmetric phase, we have,
\begin{equation}
        \langle T \eta^a_{[8]'}(x)\bar \eta^a_{[8]'}(0)\rangle
      = \langle T \eta_{[1]}(x)\bar \eta_{[1]}(0)\rangle \ , 
\end{equation}
which contain only chiral-even Dirac structures.
If all the currents couple to chiral resonances
strongly, $J={1\over 2}^\pm$ octets are
degenerate with the $J={1\over 2}^\pm$ singlets, 
which apparently is a new result\cite{shur1}. 

The chiral multiplet $(6,3)+(3,6)$ contains 
both $SU(3)_V$ octets and decuplets. Again, the currents
in the multiplet can be obtained as  
generalizations of those  
in $(1,{1\over 2})+({1\over 2},1)$ of the two-flavor case
(Eqs. (\ref{24}),(\ref{25})).
For instance,
\begin{eqnarray}
    \eta^{\mu a}_{[8]} &=& \left[\left(q^T C\sigma_{\alpha\beta} 
            q\right) \gamma_5\sigma^{\alpha\beta} 
       \gamma^\mu q\right]_{[8]}^a \ ,       \nonumber \\
    \eta^{\mu b}_{[10]} &=& \left[\left(q^T C\sigma_{\alpha\beta} q\right) 
           \gamma_5\sigma^{\alpha\beta} 
       \gamma^\mu q\right]_{[10]}^b \ . 
\end{eqnarray}
The implicit flavor indices in 10 are totally symmetric.
$\eta^{\mu b}_{[10]}$ have been used to calculate
the masses of the lowest-lying baryon decuplet in the
QCD sum rule approach \cite{yaz}.
In a chirally-symmetric phase, we have,
\begin{equation}
        \langle T \eta^{\mu a}_{[8]}(x)\bar 
   \eta^{\nu a}_{[8]}(0)\rangle
      = \langle T \eta^{\mu b}_{[10]}(x)\bar 
\eta^{\nu b}_{[10]}(0)\rangle \ , 
\end{equation}
which contain chiral-even structures only.
If these currents are dominated
by the lowest resonances, then $J={1\over 2}^\pm 
({3\over 2}^\pm)$ 
octets are degenerate with $J={1\over 2}^\pm ({3\over 2}^\pm)$ 
decuplets. 

Finally, the simplest currents in multiplet $(10,1)+(1,10)$ 
are the three-flavor generalization of
those in $({3\over 2},0)+(0,{3\over 2})$ in the
two-flavor case, 
\begin{equation}
       \eta_{[10]}^{\mu\nu b} = \left[\left(q^TC\sigma_{\alpha\beta} 
        q\right)\gamma_5\sigma^{\alpha\beta} 
        \sigma^{\mu\nu} q\right]_{[10]}^b \ , 
\end{equation}
where three implicit flavor indices are symmetric. 
An example of the chiral-singlet current 
in $(\tilde 1,1)+(1, \tilde1)$ is
\begin{equation}
       \eta_{[1]}^\mu = \left[\left(q^TC\gamma_5 
        q\right) D^\mu q - \left(q^TC
        q\right) \gamma_5 D^\mu q\right]_{[1]} \ , 
\end{equation}
which would vanish if without the covariant derivative.
In the chirally-symmetric vacuum, the two-point 
correlators of the above currents contain only chiral-even
terms.

Since there are no two simple currents in the same chiral $SU(3)$ 
representation having different $U(1)_A$ properties, 
a test of $U(1)_A$ would involve at least four baryon
currents. Correlators with three baryon currents vanish
due to the baryon number $U(1)$ symmetry. 
This generalizes the result
in Ref. \cite{birse}.

\section{comments}

In this paper, we have systematically enumerated the simplest
chiral multiplets which have meson and baryon quantum numbers.
The reason for doing this is the prospect
that QCD may have a chiral-restored phase.
If so, the chiral symmetry will 
be reflected explicitly in the correlators of hadronic currents, 
as we mentioned earlier in the Introduction.

Of course, many of the results we have stated are well known.
We repeat them here so that the reader can clearly see a 
group-theoretical organization. However, we have 
also found a number of new results which we 
summarize here:

\begin{itemize}
\item{A test of unbroken $U(1)_A$ symmetry in the two massless
flavor case is that the correlators of $\rho$ and $b_1$ currents in
parity-conjugating $({1\over 2}, {1\over 2})+({1\over 2}, {1\over 2})$
multiplets are equal.}
\item{A test of unbroken $SU(2)$ chiral symmetry is the equality of
$\rho$ and $h_1$ types of correlators and $b_1$ and $f_1$ 
types of correlators in $({1\over 2}, {1\over 2})+({1\over 2}, {1\over 2})$. This result has 
a three-flavor generalization.}
\item{We found several new interpolating currents for baryons:
the currents for $I={3\over 2}$ baryons in 
$({3\over 2}, 0)+(0, {3\over 2})$ multiplet and 
its three-flavor generalization, and the current
for singlet $\Lambda$ in $(\tilde 1,1)+(1, \tilde 1)$.}
\item{A test of unbroken $SU(2)$ chiral symmetry in baryon sector
is the equality of the $I={1\over 2}$ (nucleon) and $I={3\over 2}/2$ ($\Delta$)
two-point current correlators
in $(1, {1\over 2}) + (1, {1\over 2})$.

 $(1,1/2)+(1/2,1)$. Besides a generalization of this result
to the three flavor case, we found that the singlet $\Lambda$
and the baryon-octet correlators in $(\bar 3, 3)+(3,\bar 3)$
are equal.}
\end{itemize}

Finally, the chiral multiplets are a useful way to organize
baryon interpolating currents, which are closely related
to independent Bethe-Salpeter amplitudes. Thus, we expect
the present work to be useful also 
for the study of hadron structure. 

\acknowledgements
This work is supported in part by funds provided by the
U.S.  Department of Energy (D.O.E.) under cooperative agreement
DOE-FG02-93ER-40762.


\begin{references}
\frenchspacing

\bibitem{shur1}
E. Shuryak, Summary talk at RHIC Summer Studies, Brookhaven, July, 1996,
Hep-ph/9609249. 

\bibitem{u1a1}
E. Shuryak, Comm. Nucl. Part. Phys. {\bf 21}, 235 (1994).

\bibitem{u1a2}
J. Kapusta, D. Kharzeev, and L. McLerran, 
Phys. Rev. D {\bf 53}, 5028 (1996).

\bibitem{u1a3}
Z. Huang and X.-N. Wang, Phys. Rev. D {\bf 53}, 5034 (1996).

\bibitem{u1a}
S. Adler, Phys. Rev. {\bf 177}, 2426 (1969); J. S. Bell and R. Jackiw, 
Nuovo Cimento {\bf 51}, 47 (1969).

\bibitem{shaf}
T. Schafer and E. V. Shuryak, Phys. Rev. D {\bf 54}, 1099 (1996);
Phys. Lett. {\bf  B356}, 147 (1995).

\bibitem{BCM2}
M. C. Birse, T. D.  Cohen, J. A.  McGovern, in preparation

\bibitem{columbia}
S. Chandrasekharan and N. Christ, Nucl. Phys. B (PS) {\bf 47}, 527 (1996).

\bibitem{milc}
C. Bernard et al., hep-lat/9608026, hep-lat/9611031.

\bibitem{eta1} 
S. H. Lee and T. Hatsuda, Phys. Rev. D {\bf 54}, 1871 (1996).

\bibitem{eta2}
N. Evans, S. D. H. Hsu and M. Schwetz, Phys. Lett. B {\bf 375}, 262 (1996).

\bibitem{birse}
M. C. Birse, T. D.  Cohen, J. A.  McGovern, Phys. Lett. B {\bf 388}. 137 (1996).

\bibitem{ioffe}
B. L. Ioffe, Nucl. Phys. {\bf B188}, 317 (1981), (E) {\bf B191}, 591 (1981).

\bibitem{oka}
D. Jido, H. Kodama and M. Oka, Phys. Rev. D {\bf 54}, 4532 (1996);
D. Jido and M. Oka, hep-ph/9611322, 1996. 

\bibitem{close}
F. Close, {\it An Introduction to Quarks and Partons}
(Academic Press Inc., London, 1979).

\bibitem{yaz}
L. J. Reiders, H. R. Rubistein, S. Yazaki, Phys. Lett. {\bf 120B}, 209 (1983).

\nonfrenchspacing
\end{references}
\end{document}